\documentstyle[epsfig]{aipproc}

\begin{document}

\title{ Pressure imbalance of FRII radio source lobes: a role of
energetic proton population }

\author{M. Ostrowski$^1$ and M. Sikora$^2$}
\address{$^1$ Obserwatorium Astronomiczne, Uniwersytet Jagiello\'nski,
                   ul. Orla 171, 30-244 Krak\'ow, Poland \\
$^2$ Centrum Astronomiczne im. Miko{\l}aja Kopernika, ul. Bartycka 18,
00-716 Warszawa, Poland}

%\lefthead{LEFT head}
%\righthead{RIGHT head}
\maketitle

\begin{abstract}

Recently Hardcastle \& Worrall (2000) analyzed 63 FRII radio galaxies
imbedded in the X-ray radiating gas in galaxy clusters and concluded,
that pressures inside its lobes seem to be a factor of a few lower than
in the surrounding gas. One of explanations of the existing `blown up'
radio lobes is the existence of invisible internal pressure component
due to energetic cosmic ray nuclei (protons). Here we discuss a possible
mechanism providing these particles in the acceleration processes acting
at side boundaries of relativistic jets. The process can accelerate
particles to ultra high energies with possibly a very hard spectrum. Its
action provides also an additional viscous jet breaking mechanism. The
work is still in progress.

\end{abstract}

\section{Particle acceleration at the jet boundary}

For particles with sufficiently high energies the transition layer
between the jet and the ambient medium can be approximated as a surface
of discontinuous velocity change, a tangential discontinuity (`td'). If
particles' gyroradia (or mean free paths normal to the jet boundary) are
comparable to the actual thickness of this shear-layer interface it
becomes an efficient cosmic ray acceleration site provided the
considered velocity difference, $U$, is relativistic and the sufficient
amount of turbulence is present in the medium. The problem was
extensively discussed in early eighties by Berezhko with collaborators
(see the review by Berezhko 1990) and in the diffusive limit by Earl et
al. (1988) and Jokipii et al. (1989). The case of a relativistic jet
velocity was considered by Ostrowski (1990, 1998, 2000). The simulations
(Ostrowski 1990, cf. Bednarz \& Ostrowski 1996 for shock acceleration)
show that in favorable conditions the acceleration process acting at
relativistic tangential discontinuity of the velocity field can be very
rapid, with the time scale

$$\tau_{td} = \alpha \, {r_g \over c} \qquad , \eqno(1)$$

\noindent
where $r_g$ is a particle gyroradius in the ambient medium and -- for
efficient particle scattering -- the numerical factor $\alpha$ can be as
small as $\sim 10$. The introduced acceleration time is coupled to the
`acceleration length' $l_{td} \sim \alpha r_g$ due to particle advection
in the jet flow. In the case of a non-relativistic jet or a small
velocity gradient in the boundary shear layer the acceleration process
is of the second-order in velocity and a rather slow one. Then, the
ordinary second-order Fermi process in the turbulent medium can play a
significant, or even a dominant role in the acceleration. The
acceleration time scales can be evaluated only approximately for these
processes, and - for particles residing within the considered layer - we
can give an acceleration time scale estimate

\begin{figure}[b]
\vspace{7cm}
\includegraphics{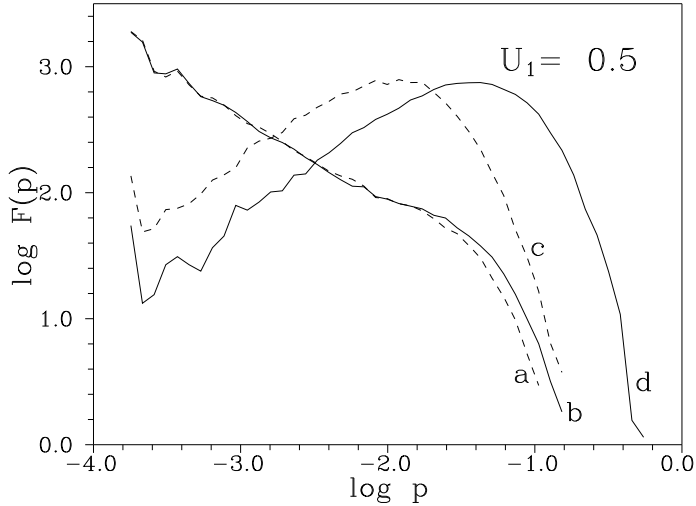}
\caption{Comparison of the escaping particle spectra formed with
wide (full lines `b' and `d') and narrow (dashed lines `a' and `c' )
turbulent cocoon surrounding the jet (cf. Ostrowski 1998). The results are
presented for two possible particle injection sites: at the terminal
shock (cases `a' and `b'), and far upstream the shock (cases `c' and
`d'), where only the boundary acceleration is effective. Spectra of
escaping particles are presented, the ones diffusively escaping from the
cocoon (mostly in `c' and `d') and these advected downstream the
terminal shock (mostly in `a' and `b'). Particle momentum unit is chosen
in a way to give its gyroradius equal the jet radius at $p = 1$.
Initial spectrum fluctuations appear near the injection momentum.}
\end{figure}

$$\tau_{II} = {r_g \over c} \, {c^2 \over  V^2 + \left( \lambda U \over
D \right)^2} \qquad , \eqno(2)$$

\noindent
where $V$ is a turbulence velocity ($\sim$ the Alfv\'en velocity for
subsonic turbulence), $\lambda$ is a mean particle free path normal to
the jet axis and $D$ is a shear layer thickness. The first term in the
denominator represents the second-order Fermi process, while the second
term is for the viscous cosmic ray acceleration. One expects that the
first term dominates at low particle energies, while the second at
larger energies, with $\tau_{II}$ approaching the value given in Eq. (1)
for $\lambda \sim D$ and $U \sim c$.

\section{Discussion}

Conditions within the large scale jets of FRII radio sources allow for
acceleration of cosmic ray protons up to energies $\sim 10^{19}$ eV
(e.g. Rachen \& Biermann 1993, Ostrowski 1998). A characteristic feature
of the boundary acceleration process in a simple considered model is
formation of very flat spectrum of {\it escaping particles}. It is due
to the on average parallel magnetic field configuration within the shear
layer, limiting the low cosmic rays escape. Such particles residing
within the shear layer volume can more efficiently stream to higher
energies due to acting the acceleration process than to escape diffusively
off the jet. At some higher energies escape becomes substantial,
leading to the spectrum cut-off formation. In the considered conditions
radiative losses are insignificant for protons.

Acting of the above mentioned processes can have pronounced consequences
for the jet propagation if the seed particle injection at `low energies'
is efficient. Accelerated particles provide a viscous agent slowing down
the jet movement. Because the jet energy is transmitted mostly to high
energy nuclei, this dissipative process can occur without significant
radiative effects. If a jet appearing from the central source with the
Lorentz factor $\sim$ a few slows down to mildly relativistic velocities
at large distances, the dissipated jet kinetic energy can be several
times larger than the one available in terminal shocks. This amount is
sufficient to explain additional pressure component providing stability
of radio lobes against pressure of the X-ray emitting gas.

If the above interpretation is true, then some further consequences of
the discussed accelerating process may arise. In particular acceleration of
nuclei could be energetically inefficient in purely electron-positron
jets, leading to less efficient jet breaking at large scales. Also, if the
considered energetic particles escape from the radio lobes too fast, the
required internal pressure component could not be sustained. A series of
such problems are under study now.

The work was supported by the {\it Komitet Bada\'n Naukowych} through
the grant PB 258/P03/99/17 (MO) and 2 P03D 00415 (MS).

\end{document}